\documentstyle[12pt]{article}
\catcode`\@=11
\def\section{\@startsection {section}{1}{0pt}{-3.5ex plus -1ex minus
 -.2ex}{2.3ex plus .2ex}{\raggedright\large\bf}}
\catcode`\@=12
\def\R{{\rm I\!R}}
\def\N{{\rm I\!N}}

\def\Q{{\mathchoice
 {\setbox0=\hbox{$\displaystyle\rm Q$}\hbox{\raise 0.15\ht0\hbox to0pt
 {\kern0.4\wd0\vrule height0.8\ht0\hss}\box0}}
 {\setbox0=\hbox{$\textstyle\rm Q$}\hbox{\raise 0.15\ht0\hbox to0pt
 {\kern0.4\wd0\vrule height0.8\ht0\hss}\box0}}
 {\setbox0=\hbox{$\scriptstyle\rm Q$}\hbox{\raise 0.15\ht0\hbox to0pt
 {\kern0.4\wd0\vrule height0.7\ht0\hss}\box0}}
 {\setbox0=\hbox{$\scriptscriptstyle\rm Q$}
 \hbox{\raise 0.15\ht0\hbox to0pt
 {\kern0.4\wd0\vrule height0.7\ht0\hss}\box0}}}}
\def\C{{\mathchoice
 {\setbox0=\hbox{$\displaystyle\rm C$}\hbox{\hbox to0pt
 {\kern0.4\wd0\vrule height0.9\ht0\hss}\box0}}
 {\setbox0=\hbox{$\textstyle\rm C$}\hbox{\hbox to0pt
 {\kern0.4\wd0\vrule height0.9\ht0\hss}\box0}}
 {\setbox0=\hbox{$\scriptstyle\rm C$}\hbox{\hbox to0pt
 {\kern0.4\wd0\vrule height0.9\ht0\hss}\box0}}
 {\setbox0=\hbox{$\scriptscriptstyle\rm C$}\hbox{\hbox to0pt
 {\kern0.4\wd0\vrule height0.9\ht0\hss}\box0}}}}
\font\fivesans=cmr5
\font\sevensans=cmr7
\font\tensans=cmr10
\newfam\sansfam
\textfont\sansfam=\tensans\scriptfont\sansfam=
 \sevensans\scriptscriptfont
\sansfam=\fivesans
\def\sans{\fam\sansfam\tensans}
\def\Z{{\mathchoice
 {\hbox{$\sans\textstyle Z\kern-0.4em Z$}}
 {\hbox{$\sans\textstyle Z\kern-0.4em Z$}}
 {\hbox{$\sans\scriptstyle Z\kern-0.3em Z$}}
 {\hbox{$\sans\scriptscriptstyle Z\kern-0.2em Z$}}}}
\newcommand{\beq}{\begin{equation}}
\newcommand{\eeq}{\end{equation}}
\newcommand{\bea}{\begin{eqnarray}}
\newcommand{\eea}{\end{eqnarray}}

\newcommand{\RG}{{\cal R}}
\newcommand{\LRG}{{\rm D}_{V_*}{\cal R}}
\newcommand{\SG}{{\cal S}}
\newcommand{\dmG}{{\rm d}\mu_\Gamma}
\newcommand{\OBS}{{\cal O}}
\newcommand{\TG}{{\cal T}}
\newcommand{\PF}{{\rm P}_{4,0}}
\newcommand{\PT}{{\rm P}_{2,2}}
\begin{document}
\begin{titlepage}
\begin{center}
{\Large The renormalized $\phi^4_4$-trajectory by \\[2mm]
perturbation theory in the running coupling} \\[10mm]
\end{center}
\begin{center}
{\Large C. Wieczerkowski} \\[10mm]
\end{center}
\begin{center}
Institut f\"ur Theoretische Physik I,
Universit\"at M\"unster, \\
Wilhelm-Klemm-Stra\ss e 9, D-48149 M\"unster, \\
wieczer@yukawa.uni-muenster.de \\[2mm]
\end{center}
\vspace{-10cm}
\hfill MS-TP1-95-06
\vspace{11cm}
\begin{abstract}
We compute the renormalized trajectory of $\phi^4_4$-theory
by perturbation theory in a running coupling. We introduce
an iterative scheme without reference to a bare action. The
expansion is proved to be finite to every order of perturbation
theory.
\end{abstract}
\end{titlepage}
%
%====================================================================
\section{Introduction}

In Wilson's renormalization group \cite{W71,WK74}
ultraviolet and infrared limit stand for
an infinite iteration of block spin transformations.
Consider for instance the ultraviolet limit of an
asymptotically free model at weak coupling. There the point
is to keep couplings under control which grow under a
block spin transformation. Such couplings are called
relevant. In weakly coupled models they can be identified
by power counting. Renormalization of a bare action
amounts to sending it through an increasing number of
block spin transformations. The image is the renormalized
action. For this limit to exist the bare couplings need to
be tuned as the number of block spin transformations is
increased. This renormalization scheme has been beautifully
implemented both within and beyond perturbation theory.
Let us mention the work of Gallavotti \cite{G85},
Gawedzki and Kupiainen \cite{GK77,GK84}, and Polchinski
\cite{P84} as a guide to the extensive literature. The
renormalized actions are located on a low dimensional curve,
parametrized by renormalized couplings. In the case of
$\phi^4$-theory we can consider a one-dimensional subspace
and speak of a renormalized trajectory. The idea is to arrange
the renormalization group flow generated from the bare action
as to converge to this renormalized trajectory in the process of
renormalization. The renormalized trajectory of an asymptotically
free model is pictured as unstable manifold of a trivial fixed
point.

Although this picture has been behind block spin
renormalization since the very beginning it has not yet
been formalized to an approach free of bare action. This
paper is a contribution to fill this gap. It extends the
analysis begun in \cite{WX94,RW95} for the hierarchical
approximation to the $\phi^4$-model with discrete momentum
space renormalization group. The renormalized trajectory
is here defined as a curve in the space of effective actions
which passes through the trivial fixed point and whose
tangent at the trivial fixed point is a $\phi^4$-vertex.
The dynamical principle which proves to be sufficiently
strong to determine this curve at least to all orders of
perturbation theory is stability under the renormalization
group. With stability we mean here that the curve is left
invariant as a set in the space of effective actions under
a block spin transformation. A renormalized action always
comes together with a sequence of descendants generated
by further block spin transformations. Even in the case of
a discrete renormalization group this sequence proves to
consist of points on a continuous curve in the space of
effective actions which is stable under a block spin
transformation. It is the computation of this curve in a
vicinity of the trivial fixed point we address.

Given a block spin transformation, we may distinguish between
the following different renormalization problems. The first
problem is an initial value problem, the analysis of the
renormalization group flow started from a particular bare
action. The second problem is a mixed boundary value problem,
where the relevant parameters are prescribed on a lower
scale, the irrelevant parameters on a higher one.
The first problem is appropriate for the infrared limit of
a Euclidean field. The second is appropriate for its ultraviolet
limit. The third problem is the question of fixed points.
In this paper we consider a generalization of the fixed point
problem. We will look for interactions which remain invariant
up to a (one dimensional) flow of a coupling parameter.
A requirement of finiteness will substitute for boundary
data. Of course the problems are interrelated. In particular
the mixed boundary value problem is a method to obtain a solution
of our generalized fixed point problem in a scaling limit.

The result is an iterative form of perturbation theory in a
running coupling. Its closest relative in the literature is
the renormalized tree expansion of Gallavotti and collaborators
\cite{G85,GN85}. Our expansion will however not be organized
in trees. Furthermore, it will from the very beginnining be free
of divergencies piled up in standard perturbation theory.
Surprisingly it will allow to treat relevant and irrelevant
couplings on the same footing. It will involve neither bare
couplings nor renormalization conditions in the original sense.
The expansion will be presented for the $\phi^4$-trajectory in
four dimensions. Most of the analysis, in fact everything except
for the treatment of the wave function term, works in arbitrary
dimensions. We therfore leave $D$ as a parameter in the equations.
Thereby dimension dependence of scale factors is exhibited.
The three dimensional case requires a modification which will be
explained elsewhere. See \cite{RW95} for a treatment of its
hierarchical approximation.\footnote{In three dimensions
the expansion parameter has a non zero scaling dimension.
As a consequence the flow of the mass term shows up a second
order correction proportional to the logarithm of the running
coupling. This type of corrections requires of a double expansion
in the running coupling and its logarithm.}

To make contact with the physical world one has to supply one
more piece of information. One has to assign a scale to a point on
the renormalized trajectory. In the presentation of this paper we
will maintain a unit scale throughout the computation.
In view of asymptotic freedom the four dimensional trajectory is
suited for the infrared limit of massless $\phi^4$-theory at
positive coupling. It is also suited for the ultraviolet limit
of massless $\phi^4$-theory at negative coupling, as been promoted
by Gawedzki and Kupiainen \cite{GK85}. In this paper we will
restrict our attention to the effective action. Its relation
with Schwinger functions is discussed for instance in \cite{BG95}.
The iterative solution of our equations can also be viewed as an
improvement program organized in powers of a running coupling.
%
%====================================================================
\section{Renormalization group}

The below analysis will be done in terms of a discrete
momentum space renormalization group transformation. A number
of applications of which is discussed for instance in recent
lectures \cite{BG95} by Benfatto and Gallavotti.
Let us consider the following discrete block spin
transformation $\RG$ on some space of interactions $V(\phi)$
of a real scalar field $\phi$ on Euclidean space $\R^D$.
Let $\RG$ be composed of a Gaussian fluctuation integral with
covariance $\Gamma$ and mean $\psi$ with a dilatation $\SG$
of the background field $\psi$. Let the fluctuation covariance
be defined by
\beq
\widehat\Gamma(p)=\frac1{p^2}\left(\widehat\chi(p)-
\widehat\chi(Lp)\right),
\label{covariance}
\eeq
where $\widehat\chi(p)$ is a momentum space cutoff function.
Its purpose is to make $\widehat\Gamma(p)$ decrease fast
outside a momentum slice $L^{-1}<|p|<1$ set by a scale
parameter $L>1$. A convenient choice is the exponential
cutoff
\beq
\widehat\chi(p)=e^{-p^2}.
\label{cutoff}
\eeq
It will be used in the following. Other choices however work
as well, for instance Pauli-Villars regularization. Then
(\ref{covariance}) defines a positive operator on the subspace
of $L_2(\R^D)$ consisting of functions $f(x)$ with
zero mode $\widehat f(0)=0$. Let $\dmG(\zeta)$ be the
associated Gaussian measure on field space. Recall its basic
property
\beq
\int\dmG(\zeta) e^{(\zeta,f)}=e^{\frac12 (f,\Gamma f)}
\label{gauss}
\eeq
and consult Glimm and Jaffe \cite{GJ87} for further information.
Let the fluctuation integral then be given by the average of
the Boltzmann factor $Z(\phi)=\exp (-V(\phi))$ with respect
to $\dmG (\zeta)$, shifted by an external background field
$\psi$. Let us introduce the notation
\beq
\left\langle Z\right\rangle_{\Gamma,\psi}=\int\dmG (\zeta)
Z(\psi+\zeta)
\label{integration}
\eeq
for this average. We can think of the momentum slice
$L^{-1}<|p|<1$ as a portion of momentum space degrees of
freedom which is integrated out. The integration of
another portion is prepared for by a dilatation of the
background field. Let this dilatation be given by
\beq
\SG\psi (x)=L^{1-\frac D2}\psi\left(\frac xL\right).
\label{dilatation}
\eeq
The exponent $\sigma=1-D/2$ is the scaling dimension of
a free massless scalar field. Anomalous rescaling will not be
considered here. Non-anomalous rescaling applies (at least) to
small perturbations of a free massless field.
The renormalization group transformation
is then defined by (\ref{integration}) composed with
(\ref{dilatation}). The following analysis will be done in
terms of the potential $V(\phi)$. The method will be perturbation
theory. The matter of stability bounds on $Z(\phi)$ will not
be addressed here. The renormalization group transformation for
the potential then reads
\beq
\RG V(\psi)= -\log\left(\left\langle\exp (-V)
\right\rangle_{\Gamma,\SG\psi}\right).
\label{rgt}
\eeq
We will restrict our attention to even potentials
$V(-\phi)=V(\phi)$.
The transformation (\ref{rgt}) preserves this property.
Potentials differing by a field independent constant will be
identified. $V(\phi)$ can for instance be normalized such
that $V(0)=0$. To maintain normalization, (\ref{rgt}) then
should be supplemented by a subtraction of $\RG V(0)$.
Technically this constant is proportional to the volume,
infinite in infinite volume. Eq. (\ref{rgt}) therefore requires
an intermediate volume cutoff to make sense. We will wipe
this technicality under the carpet, keep (\ref{rgt}) as it
stands, and ignore field independent constants.
When we perform a renormalization group transformation we will
speak of $V(\phi)$ as bare and of $\RG V(\phi)$ as effective
or renormalized potential. It should however be kept in
mind that only degrees of freedom in one momentum slice
are integrated out in a single renormalization group step.
$\zeta$ will be called fluctuation field and $\psi$ background
or block spin field. The term potential is sometimes reserved
for local interactions. Here potential will be used
synonymous with full interaction including nonlocal
interactions generated by the renormalization group.
The block scale $L$ will be kept fixed in the following.
It should not be confused with a full momentum space
cutoff of a Euclidean field. A typical value of $L$ is two.
A full cutoff could be an $N$'th power of $L$.
%
%====================================================================
\section{Trivial fixed point}

The renormalization group transformation (\ref{rgt}) has a
trivial fixed point $V_*(\phi)=0$. This fixed point is the
free massless scalar field. Eq. (\ref{rgt}) has in fact
been designed upon a momentum space decomposition of a
free massless field. The linearized renormalization
group transformation at this fixed point is given by a
Gaussian expectation value
\beq
\LRG\OBS (\psi)=\left\langle\OBS\right\rangle_{\Gamma,\SG\psi},
\label{lrgt}
\eeq
shifted by a rescaled external field. It is diagonalizable.
The eigenvectors are normal ordered
products. We will represent the potential in terms of
normal ordered products. Let us therefore briefly recall some
basic facts about normal ordering. Normal ordered products
with normal ordering covariance $v$ are generated by
\beq
:e^{(\phi,f)}:_v=e^{(\phi,f)-\frac12 (f,vf)}.
\label{normal}
\eeq
The linearized renormalization group for this generating
function is an exercise in Gaussian integration
(\ref{gauss}). The result is
\beq
\int\dmG (\zeta) :e^{(\SG\psi+\zeta,f)}:_v=
:e^{(\psi,\SG^T f)}:_{\TG v}.
\eeq
The generating function (\ref{normal}) is preserved up to a
(transposed) dilatation
\bea
\SG^T f(x)=L^{1+D/2}f(Lx)
\eea
of the source and a linear transformation
\bea
\TG v=\SG^{-1}(v-\Gamma)(\SG^{-1})^T
\label{coflow}
\eea
of the normal ordering covariance.
This linear transformation generates a flow of normal ordering.
It can be thought of as a residual renormalization group flow
taking place besides more interesting dynamical effects. It
has a line of fixed points
\beq
\widehat v(p)=\frac 1{p^2}\left(\widehat\chi(p)-C\right)
\eeq
parametrized by $C$. Let us select the point $C=0$ as normal
ordering covariance in the following, a massless covariance
with unit ultraviolet cutoff. Since it remains invariant under
(\ref{coflow}) it can be safely suppressed in the notation.
It follows that
\beq
\OBS (\phi)=\int {\rm d}^Dx_1\cdots {\rm d}^Dx_n
:\phi(x_1)^{m_1}\cdots\phi(x_n)^{m_n}:
\label{monomial}
\eeq
is an eigenvector of the linearized renormalization group
(\ref{lrgt}) with eigenvalue $L^\sigma$. The exponent is
$\sigma=nD+(m_1+\cdots+m_n)(1-D/2)$, the scaling dimension
of (\ref{monomial}). A prominent member of this family is
the $\phi^4$-vertex
\beq
\OBS(\phi)=\int{\rm d}^Dx:\phi(x)^4:
\label{vertex}
\eeq
with scaling dimension $4-D$. It is therefore called relevant in
$D<4$, marginal in $D=4$, and irrelevant in $D>4$ dimensions.
General eigenvectors are given by homogeneous kernels in
real or momentum space and also involve derivatives of fields.
See the review \cite{W76} by Wegener.
%
%====================================================================
\section{Perturbation theory}

In a vicinity of the trivial fixed point the transformation (\ref{rgt})
can be computed by means of perturbation theory. The perturbation
expansion for the effective potential reads
\beq
\RG V(\psi)=\sum_{n=1}^\infty\frac{(-1)^{n+1}}{n!}
\left\langle [V;]^n\right\rangle^T_{\Gamma,\SG\psi}.
\label{cumulant}
\eeq
The superscript $^T$ indicates truncated expectation values.
Truncated expectation values are defined by
\beq
\left\langle\prod_{i=1}^n[\OBS_i;]\right\rangle^T=
\left[
\left(\prod_{i=1}^n\frac\partial{\partial\lambda_i}
\right)
\log\left\langle\exp\left(\sum_{i=1}^n\lambda_i
\OBS_i\right)\right\rangle
\right]_{\lambda_1=\cdots=\lambda_n=0}.
\eeq
Notice that the truncated expectation values
$\left\langle [V;]^n\right\rangle^T_{\Gamma,\SG\psi}$
are the cumulants associated with the moments
$\left\langle V^n\right\rangle_{\Gamma,\SG\psi}$.
The perturbation expansion (\ref{cumulant}) is
therefore also known as cumulant expansion.
Let us consider the situation when the potential comes
in form of a power series
\beq
V(\phi|g)=\sum_{n=1}^\infty\frac{g^n}{n!}V^{(n)}(\phi)
\label{potential}
\eeq
of a coupling parameter $g$. The zeroth order interaction
will always be $V^{(0)}(\phi)=0$ in the following.
In the bare perturbation expansion
for a single renormalization group transformation
the effective interaction is expanded again in the
bare coupling. Let us introduce the notation
\beq
\RG V(\psi|g)=\sum_{n=1}^\infty\frac{g^n}{n!}
(\RG V)^{(n)}(\psi)
\label{bare}
\eeq
for this expansion. The individual orders of
perturbation theory are given by sums of
truncated expectation values
\beq
(\RG V)^{(n)}(\psi)=\sum_{m=1}^n\frac{(-1)^{m+1}}{m!}
\sum_{\begin{array}{c}l_1,\dots,l_m\in\N\\
l_1+\cdots+l_m=n\end{array}}\left(\begin{array}{c}
n\\l_1\cdots l_{m-1}\end{array}\right)
\left\langle V^{(l_1)};\dots;V^{(l_m)}
\right\rangle^T_{\Gamma,\SG\psi}.
\label{orders}
\eeq
The sum over $(l_1,\dots,l_m)$ is restricted to $m$-tupels
in $\{1,\dots,n\}^m$ such that $l_1+\cdots+l_m=n$ and is finite.
The multinomial coefficient is given by
\beq
\left(\begin{array}{c}n\\l_1\cdots l_{m-1}\end{array}\right)
=\frac{n!}{\prod_{k=1}^m l_k!}
\eeq
with $l_m=n-l_1-\cdots l_{m-1}$.
To lowest orders we find the following
explicit expressions for the effective interactions
\bea
(\RG V)^{(1)}(\phi)&=&\left\langle V^{(1)}
\right\rangle_{\Gamma,\SG\psi},\label{rg1}\\
(\RG V)^{(2)}(\phi)&=&\left\langle V^{(2)}
\right\rangle_{\Gamma,\SG\psi}-
\left\langle V^{(1)};V^{(1)}
\right\rangle^T_{\Gamma,\SG\psi},\label{rg2}\\
(\RG V)^{(3)}(\phi)&=&\left\langle V^{(3)}
\right\rangle_{\Gamma,\SG\psi}-3
\left\langle V^{(2)};V^{(1)}
\right\rangle^T_{\Gamma,\SG\psi}+
\left\langle V^{(1)};V^{(1)};V^{(1)}
\right\rangle^T_{\Gamma,\SG\psi}.\label{rg3}
\eea
Let us have a closer look at two contributions to
(\ref{orders}). The highest order bare interaction
appearing to order $n$ is $V^{(n)}(\phi)$. It
contributes a term
$\left\langle V^{(n)}\right\rangle_{\Gamma,\SG\psi}$.
It is thus transformed according to the linearized
renormalization group (\ref{lrgt}).
The perturbative corrections to the linearized
renormalization group depend on lower orders
$V^{(m)}(\phi)$ with $1\leq m\leq n-1$ only.
The first order $V^{(1)}(\phi)$ contributes by itself
a term $(-1)^{n+1}\left\langle [V^{(1)};]^n
\right\rangle^T_{\Gamma,\SG\psi}$ to $(\RG V)^{(n)}(\psi)$.
These interactions need to be carried along immediately
to order $n$ when some first order interaction
enters the game. In a minimal scheme no further
interactions would be introduced to this order.

(\ref{orders}) leaves us with an expansion for the
effective potential in terms of the bare coupling.
This expansion is not appropriate for an iteration of
renormalization group transformations. The appropriate
expansion is an expansion in powers of the effective
coupling. This requires a reorganization of (\ref{orders})
in powers of the effective coupling. Let this effective
coupling come as a power series
\beq
\beta (g)=\sum_{n=1}^\infty
\frac{g^n}{n!} b_n
\label{effective}
\eeq
in the bare coupling.
The function $\beta (g)$ is not a Callan-Symanzik $\beta$-function
in a literal sense but its block spin analogue.
The inverse reorganization goes as
follows. Let the effective potential be given as a power
series (\ref{potential}) of the effective coupling. Then we
can substitute (\ref{effective}) and expand the result in
powers
\beq
V(\psi|\beta(g))=\sum_{n=1}^\infty
\frac{g^n}{n!}(V\circ\beta)^{(n)}(\psi)
\label{organization}
\eeq
of the bare coupling. The coefficients in this reorganized
expansion (\ref{organization}) are given by
\beq
(V\circ\beta)^{(n)}(\psi)=\sum_{m=1}^n\frac1{m!}
\sum_{\begin{array}{c}l_1,\dots,l_m\in\N\\
l_1+\cdots+l_m=n\end{array}}\left(\begin{array}{c}
n\\l_1\cdots l_{m-1}\end{array}\right)
b_{l_1}\cdots b_{l_m} V^{(m)}(\psi).
\label{expanded}
\eeq
in terms of the coefficients of (\ref{potential}) and
(\ref{effective}). The zeroth order of (\ref{effective})
is put to $b_0=0$. To lowest orders (\ref{expanded}) is
given by
\bea
(V\circ\beta)^{(1)}(\psi)&=&b_1 V^{(1)}(\psi),\label{be1}\\
(V\circ\beta)^{(2)}(\psi)&=&b_2 V^{(1)}(\psi)+
(b_1)^2 V^{(2)}(\psi),\label{be2}\\
(V\circ\beta)^{(3)}(\psi)&=&b_3 V^{(1)}(\psi)+
3 b_1 b_2 V^{(2)}(\psi)+(b_1)^3 V^{(3)}(\psi).\label{be3}
\eea
This combinatorial exercise completes the setup for
perturbation theory.

If we would iterate (\ref{orders}) we would
end up with the tree expansion of Gallavotti \cite{G85}.
See \cite{FHRW88} for a pedagogical account of this well
developed technology. We will not organize perturbation
theory in terms of trees. Also we will not start with a
bare perturbation expansion but directly attack the
renormalized series. Let us conclude this section with the
remark that the effective potential (\ref{cumulant})
is the generating function of free propagator amputated
connected Green's functions with vertices (\ref{potential})
and propagator $\Gamma$. The terms in (\ref{orders})
can therefore be expressed in terms of connected
Feynman diagrams.
%
%====================================================================
\section{$\phi^4$-trajectory}

Let us pose the following renormalization problem. We seek
a potential $V(\phi|g)$ depending on a running coupling $g$
with the following properties. ("running" will be explained
below.)
\\[2mm]{\sl
1) $V(\phi|g)$ is a power series
\beq
V(\phi|g)=\sum_{n=1}^\infty\frac{g^n}{n!}V^{(n)}(\phi)
\label{power}
\eeq
in the coupling parameter $g$.
}\\[2mm]
The zeroth order is here $V^{(0)}(\phi)=0$. We will treat
of (\ref{power}) as a formal perturbation of the trivial
fixed point. The important question of summability of
(\ref{power}) will not be addressed.
\\[2mm]{\sl
2) The first order in (\ref{power}) is a $\phi^4$-vertex
\beq
V^{(1)}(\phi)=\frac1{4!}\int{\rm d}^Dx:\phi(x)^4:.
\label{one}
\eeq
The $n$'th order in (\ref{power}) is a polynomial
\beq
V^{(n)}(\phi)=\sum_{m=1}^{n+1}\frac1{2m!}
\int {\rm d}x_1\dots{\rm d}x_{2m}
V^{(n)}_{2m}(x_1,\dots,x_{2m})
:\phi(x_1)\cdots\phi(x_{2m}):
\label{polynomial}
\eeq
in the field $\phi$.
}\\[2mm]
Let us imagine a bare theory consisting purely of a
first order vertex (\ref{one}), possibly shouldered by
second order mass and wave function counterterms. The
effective interactions generated in the course of its
renormalization group flow will not remain of this simple
form. The first renormalization group step already leaves
us with an infinite set of higher order vertices, among
which are for example second order mass and wave function
terms. A main theme of this paper is to add all these
higher order interactions to the bare theory from the
beginning since they will anyway be generated once
(\ref{one}) enters the theory. In a minimal scheme no
other vertices are introduced than those enforced by the
presence of (\ref{one}). We can then think of (\ref{one})
as the germ of the theory. Other trajectories emerging
from the trivial fixed point are of interest as well,
for instance the $\phi^6$-trajectory. There the first order
is a $\phi^6$-vertex. Our leitmotiv here is to keep
the effective interaction as minimal as possible.
The highest connected vertex generated from $n$ first order
$\phi^4$-vertices has $2n+2$ legs. Higher vertices will
appear at this order only if they are introduced by hand.
(\ref{polynomial}) excludes this option. This form of
potential iterates through the renormalization group.
Field independent terms are discarded. We consider even
powers of fields only.
\\[2mm]{\sl
4) The kernels in (\ref{polynomial}) are Euclidean invariant
distributions. They are given by Fourier integrals
\beq
V^{(n)}_{2m}(x_1,\dots,x_{2m})=
\int\frac{{\rm d}^Dp_1}{(2\pi)^D}\cdots
\frac{{\rm d}^Dp_{2m}}{(2\pi)^D}
e^{i\sum_{l=1}^{2m}p_lx_l}
(2\pi)^D\delta\left(\sum_{l=1}^{2m}p_l\right)
\widehat V^{(n)}_{2m}(p_1,\dots,p_{2m}).
\label{fourier}
\eeq
With the $\delta$-function removed their Fourier transforms
are symmetric ${\cal C}^\infty$-functions on momentum space
$\R^D\times\cdots\times\R^D$. They satisfy the bounds
\beq
\left\| p^\alpha\frac{\partial^{|\alpha |}}
{\partial p^\alpha}\widehat V^{(n)}_{2m}
\right\|_{\infty,\epsilon}
=\sup_{(p_1,\ldots,p_{2m})\in {\cal M}_{2n}}
\left\{\left\vert p^\alpha\frac{\partial^{|\alpha |}}
{\partial p^\alpha}\widehat V^{(n)}_{2m}
(p_1,\ldots,p_{2m})\right\vert
e^{-\epsilon (|p_1|+\ldots+|p_{2m}|)}\right\}
<\infty.
\label{supremum}
\eeq
for any $\epsilon >0$. Here ${\cal M}_{2m}$ denotes the
subset of momenta $(p_1,\ldots,p_{2m})\in\R^D\times\cdots
\times\R^D$ with $p_1+\dots +p_{2m}=0$. Furthermore,
$\alpha=(\alpha_{i,\mu})\in\N^{2mD}$ is a multi-index and
$|\alpha|=\sum_{i,\mu}\alpha_{i,\mu}$.}\\[2mm]
The kernels in (\ref{fourier}) are the unknowns in this
approach. (\ref{supremum}) is meant as requirement on what
kind of kernels we will hold look out for.
This condition of finiteness substitutes for boundary data
which is necessary for instance in \cite{G85} and \cite{P84}.
On the practical side we want to be certain that our perturbation
theory is finite to every order. The $L_{\infty,\epsilon}$-norm
in momentum space is a convenient but not the only possible
criterion. See \cite{FHRW88} for further inspiration.
The large momentum bound implied by (\ref{supremum}) is
rather wasteful. The accurate large momentum behavior is
polynomial in powers and logarithms of momenta. Their origin
is a Taylor expansion of the quadratic and quartic kernel in
four dimensions. The Taylor expansion is compelled by power
counting. The remainders are irrelevant. The norm estimates
present bounds on these irrelevant remainder terms. The Taylor
coefficients are of course also required to be finite. This is
understood when we speak of smooth functions on momentum space.
Smoothness is a little more than what is needed.
Existence of required derivatives at zero momentum
(or some general subtraction point) together with
$L_{\infty,\epsilon}$-bounds on the remainders is fully sufficient.
In the ultraviolet problem with exponential cutoff
we can afford the luxury of smoothness. Euclidean
invariance reduces the Taylor coefficients in four dimensions
to a mass term, a wave function term, and the $\phi^4$-vertex.
Note that (\ref{rgt}) respects Euclidean invariance. It is
therefore broken only if we break it by hand. Note also that
translation invariance by itself affects momentum space power
counting.
\\[2mm]{\sl
5) The four point kernel (\ref{fourier}) is
\beq
\widehat V^{(n)}_4(0,0,0,0)=\delta_{n,1}
\label{coupling}
\eeq
at zero momentum.
}\\[2mm]
The zero momentum condition (\ref{coupling}) is part of
the definition of the expansion parameter $g$. If $g$
is traded for another parameter $g^\prime(g)=g+O(g^2)$
then (\ref{coupling}) is not true anymore. There is
nothing wrong with a redefinition of $g$, although it
looks unnecessary to introduce a $\phi^4$-vertex at some
higher order when it is already present to first order.
\\[2mm]{\sl
6) There exists a power series
\beq
\beta(g)=\sum_{n=1}^\infty\frac{g^n}{n!}b_n
\label{beta}
\eeq
such that
\beq
\RG V(\psi|g)=V(\psi|\beta(g))
\label{invariance}
\eeq
to every order in $g$.
}\\[2mm]
This condition is the core of our approach. It says that
$V(\phi|g)$ remains of the same form under the
renormalization group up to flow of $g$. It is therefore
called running coupling. (\ref{invariance}) will prove
to be strong enough to determine both $V(\phi|g)$ and
$\beta(g)$ to every order in $g$. In $D=4$ dimensions
this scheme will require a modification due to wave function
renormalization. We will need to introduce a first order
wave function term. But let us postpone this modification
for the moment. A potential with the property (\ref{invariance})
will be said to scale.

It is amusing to think of the solution
$V(\phi|\cdot):g\mapsto V(\phi|g)$ as a parametrized curve
in interaction space. Anticipating a future renormalization
group geometry we can then say that: 1) $V(\phi|\cdot)$
visits the trivial fixed point at $g=0$. 2) The tangent to
$V(\phi|\cdot)$ at $g=0$ is the $\phi^4$-vertex. 3)
$V(\phi|\cdot)$ is stable under the renormalization group
as a set in interaction space. We call this curve
$\phi^4$-trajectory.
%
%====================================================================
\section{Scaling equations}

Expanding both sides of (\ref{invariance}) in powers of $g$,
we obtain a system of scaling equations for the perturbations
series (\ref{power}) and (\ref{beta}). This system is organized
into a recursion relation. From (\ref{rg1}), (\ref{rg2}),
(\ref{rg3}) and (\ref{be1}), (\ref{be2}), (\ref{be3}) we find
to lowest orders equations of the form
\bea
\left\langle V^{(1)}\right\rangle_{\Gamma,\SG\psi}-
\quad b_1 V^{(1)}(\psi)&=&0,\\
\left\langle V^{(2)}\right\rangle_{\Gamma,\SG\psi}-
(b_1)^2 V^{(2)}(\psi)&=&b_2 V^{(1)}(\psi)+
\left\langle V^{(1)};V^{(1)}\right\rangle^T_{\Gamma,\SG\psi},
\label{secord}\\
\left\langle V^{(3)}\right\rangle_{\Gamma,\SG\psi}-
(b_1)^3 V^{(3)}(\psi)&=&b_3 V^{(1)}(\psi)+3b_1 b_2 V^{(2)}(\psi)+
3\left\langle V^{(1)};V^{(2)}\right\rangle^T_{\Gamma,\SG\psi}-
\nonumber\\& &
\left\langle V^{(1)};V^{(1)};V^{(1)}\right\rangle^T_{\Gamma,\SG\psi}.
\eea
To first order, scaling requires $V^{(1)}(\psi)$ to be an
eigenvector of the linearized renormalization group, and $b_1$
to be the eigenvalue. This is the case for the $\phi^4$-vertex
with the familiar eigenvalue $b_1=L^{4-D}$. The first order equation
is special in that it is homogeneous. To higher orders we meet
inhomogeneous linear equations. Their general form for $n\geq 2$
is
\bea
&& \left\langle V^{(n)}\right\rangle_{\Gamma,\SG\psi}-
(b_1)^n V^{(n)}(\psi)=
\nonumber\\&&\quad
b_n V^{(1)}(\psi)+
(-1)^n\left\langle V^{(1)};\cdots;V^{(1)}
\right\rangle^T_{\Gamma,\SG\psi}+
\sum_{\begin{array}{c}l_1,\dots,l_m\in\N\\
l_1+\cdots+l_m=n\end{array}}\left(\begin{array}{c}
n\\l_1\cdots l_{m-1}\end{array}\right)
\nonumber\\&&\quad
\sum_{m=2}^{n-1}\frac1{m!}
\biggl(b_{l_1}\cdots b_{l_m} V^{(m)}(\psi)+
(-1)^m\left\langle V^{(l_1)};\cdots;V^{(l_m)}
\right\rangle^T_{\Gamma,\SG}\biggr).
\label{recursion}
\eea
We give the name $L^{n(4-D)}K^{(n)}(\psi)$ to the right hand
side of (\ref{recursion}). It is determined by the effective
potential $V^{(m)}(\psi)$ to lower orders $1\leq m\leq n-1$.
Eq. (\ref{recursion}) poses an inhomogeneous linear problem
for $V^{(n)}(\psi)$. Its left hand side is diagonalized by
normal ordering. Insert the expansion (\ref{polynomial}) in
normal ordered powers of fields to obtain
\bea
&&\left\langle V^{(n)}\right\rangle_{\Gamma,\SG\psi}-
(b_1)^n V^{(2)}(\psi)=
\nonumber\\&&\quad
\sum_{m=1}^{n+1}\frac1{(2m)!}\int{\rm d}^Dx_1\cdots
{\rm d}^Dx_{2m} :\psi(x_1)\cdots\psi(x_{2m}):
\nonumber\\&&\quad
\biggl(L^{m(D+2)}V^{(n)}_{2m}(Lx_1,\dots,Lx_{2m})-
L^{n(4-D)}V^{(n)}(x_1,\dots,x_{2m})\biggr).
\label{left}
\eea
The linearized renormalization group thus performs a scale
transformation of a kernel.\footnote{Homogeneous kernels
therefore give eigenvectors of the linearized transformation.}
The difference of exponents defines an order dependent real
space power counting
\beq
\sigma(m,n)=m(D+2)-n(4-D).
\eeq
The right hand side of (\ref{recursion}) can now be expanded
in the same manner. We introduce the notation
\beq
K^{(n)}(\psi)=
\sum_{m=1}^{n+1}\frac1{(2m)!}\int{\rm d}^Dx_1\cdots
{\rm d}^Dx_{2m} :\psi(x_1)\cdots\psi(x_{2m}):
K^{(n)}_{2m}(x_1,\dots,x_{2m})
\label{right}
\eeq
for this expansion in normal ordered powers of fields. From
(\ref{left}) and (\ref{right}) together we then deduce the  
equations
\beq
L^{\sigma(m,n)}V^{(n)}_{2m}(Lx_1,\dots,Lx_{2m})-
V^{(n)}_{2m}(x_1,\dots,x_{2m})=
K^{(n)}_{2m}(x_1,\dots,x_{2m}).
\label{realscale}
\eeq
The problem of renormalized perturbation theory has thus been
reduced to the solution of (\ref{realscale}). Recall that the
right hand side is determined by vertices of lower orders and
cutoff propagators. The kernels are distributions on copies of
real space. Fourier transformation turns (\ref{realscale}) into
\beq
L^{\widehat\sigma(m,n)}
\widehat V^{(n)}_{2m}(\frac{p_1}L,\dots,\frac{p_{2m}}L)-
\widehat V^{(n)}_{2m}(p_1,\dots,p_{2m})=
\widehat K^{(n)}_{2m}(p_1,\dots,p_{2m}).
\label{momscale}
\eeq
Here the $\delta$-function due to translation invariance has
been removed. Eq. (\ref{momscale}) is the main dynamical
equation in this approach. The task is to search for finite
solutions in the sense of (\ref{supremum}). In momentum space
we find an order dependent power counting
\beq
\widehat\sigma(m,n)=D-m(D-2)-n(4-D).
\label{mompower}
\eeq
In the sequel the attributes relevant, marginal, and irrelevant
will be used to distinguish whether (\ref{mompower}) is positive,
zero, or negative. Notice that (\ref{mompower}) becomes
\bea
\widehat\sigma(m,n)&=&3-m-n,\quad D=3,\\
\widehat\sigma(m,n)&=&4-2m,\quad D=4,
\eea
in three and four dimensions respectively. Thus in three
dimensions three interactions\footnote{Interaction here
refers to the full kernel and comprises its value as well
as derivatives at zero momentum.} are non-irrelevant, while
there are infinitely many in four dimensions. Labelling
interactions by a pair $(m,n)$ we have:
\begin{center}\begin{tabular}{ll|c|c}
 &(m,n)&{\rm relevant}&{\rm marginal}\\
D& & & \\ \hline
3& & (1,1) & (1,2),\quad (2,1) \\ \hline
4& & (1,n) & (2,n)
\end{tabular}\end{center}
The conclusion is of course that the theory is super-renormalizable
in three and renormalizable in four dimensions. In the BPHZ
scheme this leads to the complication that the renormalizable
case has infinitely many divergent graphs. In the present
approach (with no divergent graphs) the renormalizable case will
not be more complicated than the super-renormalizable one.

%%%%%%%%%%%%%%%%%%%%%%%%%%%%%%%%%%%%%%%%%%%%%%%%%%%%%%%%%%%%%%%%%%%%%
Consider a general difference equation of the form
\beq
L^\sigma f\left(\frac pL\right)-f(p)=g(p)
\label{simple}
\eeq
for functions on $\R^N$. Assume that $L>1$, $\sigma$ is integer
valued, and $g\in{\cal C}^\infty (\R^N)$. We distinguish two
cases, the irrelevant case with $\sigma <0$ and the relevant
case with $\sigma >0$.
\\[2mm]{\sl
Let $\sigma <0$ and $g\in{\cal C}^\infty (\R^n)$.
\\[2mm]
1) The series
\beq
f(p)=-\sum_{m=0}^\infty L^{m\sigma} g\left(\frac p{L^m}\right)
\label{iter}
\eeq
is uniformly convergent on compact subsets of $\R^N$.
It defines a function $f\in{\cal C}^\infty (\R^n)$. \\[2mm]
2) The function $f$ given by (\ref{iter}) is a solution to
(\ref{simple}). It is unique in the space of smooth functions.
\\[2mm]
3) Let $|g|_{\infty,\epsilon}<\infty$.\footnote{Recall that
$|g|_{\infty,\epsilon}=\sup_{p\in\R^N}|g(p)|\exp
(-\epsilon |p|)$.} Then $f$ satisfies the bound
\beq
|f|_{\infty,\epsilon}\leq \frac 1{1-L^\sigma}
|g|_{\infty,\epsilon}.
\eeq
If the derivatives of $g$ are $L_{\infty,\epsilon}$-bounded,
so are the derivatives of $f$.
}\\[2mm]
Formula (\ref{iter}) is obtained by iteration (\ref{simple}).
Its convergence follows from the Weierstrass condition for
uniform convergence \cite{WW86}. It is checked to satisfy
(\ref{simple}). The difference of two solutions to (\ref{simple})
satisfies the homogeneous equation
$L^\sigma f\left(\frac pL\right)-f(p)=0.$
When $\sigma$ is negative its only smooth solution is zero.
The $L_{\infty,\epsilon}$-bound is elementary.

%%%%%%%%%%%%%%%%%%%%%%%%%%%%%%%%%%%%%%%%%%%%%%%%%%%%%%%%%%%%%%%%%%%%%
Thus eq.(\ref{simple}) has been solved in the irrelevant case.
In the relevant case the trick is to reduce the degree of
relevancy by taking derivatives. Deriving (\ref{simple}) once,
we have
\beq
L^{\sigma-1} \frac{\partial f}{\partial p}\left(\frac pL\right)-
\frac{\partial f}{\partial p}(p)=\frac{\partial g}{\partial p}(p).
\label{partial}
\eeq
A partial derivative therefore reduces powercounting by one
unit. The value of the derivative at the origin is obtained
by evaluation of (\ref{partial}). The strategy is therefore
a Taylor expansion with remainder term to an order where the
remainder becomes irrelevant.
\\[2mm]{\sl
Let $\sigma\geq 0$ and $g\in{\cal C}^\infty (\R^n)$. Then
\bea
f(p)&=&\sum_{|\alpha |\leq\sigma}\frac 1{\alpha !}
\frac{\partial^{|\alpha |}f}{\partial p^\alpha}(0)
p^\alpha+\sum_{|\alpha |=\sigma+1}\frac 1{\alpha !}
\int_0^1{\rm d}t(1-t)^\sigma
\frac{\partial^{|\alpha |}f}{\partial p^\alpha}(tp)
p^\alpha
\label{taylor}
\eea
is a solution to (\ref{simple}) if and only if
\bea
\left( L^{\sigma-|\alpha |}-1\right)
\frac{\partial^{|\alpha |}f}{\partial p^\alpha}(0)&=&
\frac{\partial^{|\alpha |}g}{\partial p^\alpha}(0),
\quad |\alpha |\leq \sigma, \label{coeffs}\\
L^{\sigma-|\alpha |}
\frac{\partial^{|\alpha |} f}{\partial p^\alpha}
\left(\frac pL\right)-
\frac{\partial^{|\alpha |} f}{\partial p^\alpha}(p)&=&
\frac{\partial^{|\alpha |} g}{\partial p^\alpha}(p),
\quad |\alpha |=\sigma+1.
\label{remain}
\eea
}\\[2mm]
The proof is obvious. As a consequence smooth solutions
exist only when all marginal Taylor coefficients of $g$
vanish at the origin. The marginal Taylor coefficients
of the solutions are free parameters.
\\[2mm]{\sl
Let $\sigma\geq 0$ and $g\in{\cal C}^\infty (\R^n)$. \\[2mm]
1) Smooth solutions to (\ref{simple}) exist if and only if
\beq
\frac{\partial^{|\alpha |} g}{\partial x^\alpha}(0)=0
\label{margcon}
\eeq
for all multi-indices with $|\alpha |=\sigma$. \\[2mm]
2) In this case, we obtain a set of solutions which can be
parametrized by the Taylor coefficients
\beq
c_\alpha=\frac{\partial^{|\alpha |} f}{\partial x^\alpha}(0)
\label{tcoeffs}
\eeq
for all multi-indices with $|\alpha |=\sigma$.
}\\[2mm]
The proof is obvious. The grain of salt is that the
marginal Taylor coefficients are not determined by eq.
(\ref{simple}). The relevant Taylor coefficients and
the irrelevant Taylor remainder are however explicitly
known from (\ref{iter}), (\ref{coeffs}), and (\ref{remain}).

If the Taylor expansion (\ref{taylor}) is pushed farther
than to order $\sigma$, then also the Taylor coefficients
with negative power counting obey (\ref{coeffs}). The
iterative and the direct solution of (\ref{coeffs}) coincides
when $\sigma-|\alpha |<0$. Thus in terms of Taylor coefficients
there is surprisingly no difference between the relevant and
the irrelevant ones. Finally, remark that also the situation with
non-integer scaling dimension $\sigma$ has applications
in the renormalization group. It is possible to cook up
models with irrational scaling dimensions. Then the
scheme is particularly simple because marginal coefficients
are absent.

With this interlude we are essentially finished.
The irrelevant kernels in (\ref{momscale}) are solved
by iteration. The non-irrelevant kernels are Taylor expanded.
The Taylor coefficients are solved directly.
%
%====================================================================
\section{Marginalia}

Marginal eigenvectors require a separate treatment. The
following scheme is designed for $\phi^4$-theory in $D=4$
dimensions. More generally it applies to models whose
coupling is dimensionless. The underlying idea is that
marginal eigenvectors possess a logarithmic flow. If also
the coupling is marginal we may express the former
logarithmic flow in terms of the latter. $\phi^4$-theory
in $D=3$ dimensions for instance calls for a different
approach which will be explained elsewhere. There the
coupling flows powerlike and one needs to introduce terms
proportional to logarithms of the coupling to handle the
marginal eigenvectors. For the rest of this section we put
$D=4$. Let us first do the computation of the $\beta$-function
without wave function renormalization to explain the scheme
in a slightly simpler setting. It is convenient to write it
in terms of projectors on the marginal eigenvectors. Their
meaning is nothing but Taylor expansion in momentum space.
For a general Euclidean invariant functional
\beq
V(\phi)=\sum_{m=1}^{\infty}\frac1{2m!}
\int {\rm d}^Dx_1\dots{\rm d}^Dx_{2m}
V_{2m}(x_1,\dots,x_{2m})
:\phi(x_1)\cdots\phi(x_{2m}):
\label{functional}
\eeq
define projectors
\bea
\PF V(\phi)&=&\frac 1{4!}\int{\rm d}^Dx_1 :\phi(x_1)^4:
\int{\rm d}^Dx_2{\rm d}^Dx_3{\rm d}^Dx_4 V_4(x_1,x_2,x_3,x_4),
\label{project1}\\
\PT V(\phi)&=&\frac 12\int{\rm d}^Dx_1 :(\partial\phi(x_1))^2:
\frac{-1}D\int{\rm d}^Dx_2 (x_1-x_2)^2V_2(x_1,x_2),
\label{project2}
\eea
on the marginal eigenvectors
\bea
\OBS_{4,0}(\phi)&=&\frac 1{4!}\int{\rm d}^Dx :\phi(x)^4:,
\label{eigen1}\\
\OBS_{2,2}(\phi)&=&\frac 12\int{\rm d}^Dx :(\partial\phi(x))^2:.
\label{eigen2}
\eea
These eigenvectors correspond to the particular kernels
\bea
V_{2n}(x_1,\dots,x_{2n})&=&\delta_{n,1}\delta(x_2-x_1)
\delta(x_3-x_1)\delta(x_4-x_1),
\label{kernel1}\\
V_{2n}(x_1,\dots,x_{2n})&=&\delta_{n,1}
\frac{-\partial^2}{\partial x_1^2}\delta(x_1-x_2),
\label{kernel2}
\eea
respectively, symmetrized in their entries. It is clear that
they are reproduced by the projectors (\ref{project1}) and
(\ref{project2}). Furthermore it is clear that the projectors
satisfy
\bea
\PF\left\langle V\right\rangle_{\Gamma,\SG\psi}&=&
L^{4-D}\PF V(\psi),
\label{property1}\\
\PT\left\langle V\right\rangle_{\Gamma,\SG\psi}&=&
\PT V(\psi).
\eea
The wave function term (\ref{eigen2}) is marginal (at order
zero) in all dimensions and the $\phi^4$-vertex (\ref{eigen1})
in four dimensions. (\ref{project2}) and (\ref{project1}) are
projectors on the marginal content of (\ref{functional}).
Let us now consider the $\beta$-function. For all orders
$n\geq 2$, the definition of the coupling constant implies that
\beq
\PF V^{(n)}(\phi)=0.
\eeq
That is the higher orders do not contain an eigenvector
(\ref{eigen2}). Applying (\ref{project2}) to both sides of
(\ref{recursion}), and using (\ref{property1}) to conclude
that the left side projects to zero, we find
\bea
b_n \PF V^{(1)}(\psi)&=&\sum_{m=2}^{n-1}
\frac{(-1)^{m+1}}{m!}
\sum_{\begin{array}{c}l_1,\dots,l_m\in\N\\
l_1+\cdots+l_m=n\end{array}}\left(\begin{array}{c}
n\\l_1\cdots l_{m-1}\end{array}\right)
\nonumber\\
&&\quad\PF\left\langle V^{(l_1)};\cdots;V^{(l_m)}
\right\rangle^T_{\Gamma,\SG\psi}.
\label{betan}
\eea
The projector $\PF$ selects the contributions built from
lower order vertices, fluctuation propagators, and normal
ordering propagators to the effective $n$'th order $\phi^4$-vertex.
This equation determines the coefficient $b_n$.

Let us now also introduce a wave function renormalization. This
requires a slight modification of the renormalization scheme
used so far. The second order condition on the wave function
part is
\beq
b_2\PT V^{(1)}(\psi)=-\PT\left\langle V^{(1)};V^{(1)}
\right\rangle^T_{\Gamma,\SG\psi}.
\label{secondwave}
\eeq
From it we see that we cannot leave the first order to
remain a pure $\phi^4$-vertex (\ref{one}) unless the
right side of (\ref{secondwave}) does not contain a wave
function term. This is unfortunately not the case. A wave
function term is indeed generated to second (and arbitrary)
order of perturbation theory. Therefore we have to include
a wave function term already to first order in the running
coupling and replace (\ref{one}) by
\beq
V^{(1)}(\phi)=\frac1{4!}\int{\rm d}^Dx:\phi(x)^4:+
\frac {\zeta_1}2\int{\rm d}^Dx :(\partial\phi(x))^2:.
\label{newone}
\eeq
Here $\zeta_1$ is a new parameter, the first order wave
function renormalization. Its value is determined by the
second order equation (\ref{secondwave}). On a first
inspection this equation looks quadratic in $\zeta_1$.
This is not the case since
\beq
\PT\left\langle\OBS_{2,2};\OBS_{2,2}
\right\rangle^T_{\Gamma,\SG\psi}=0.
\label{vanish1}
\eeq
The reason is that the fluctuation propagator (\ref{covariance})
is regular at zero momentum. Another fact which is pleasantly
welcome is
\beq
\PT\left\langle\OBS_{2,2};\OBS_{4,0}
\right\rangle^T_{\Gamma,\SG\psi}=0.
\label{vanish2}
\eeq
As a consequence the first order wave function term does not
alter the second order coefficient $b_2$ of the $\beta$-function.
A wave function term on an external leg of the $\phi^4$-vertex
makes the resulting amplitude vanish at zero momentum due to
the regularity of (\ref{covariance}) at zero
momentum. As a consequence $\zeta_1$ does not appear in
\beq
b_2\PF V^{(1)}(\psi)=-\PF\left\langle V^{(1)};V^{(1)}
\right\rangle^T_{\Gamma,\SG\psi}.
\label{betatwo}
\eeq
from which $b_2$ is computed. Thus we can first find $b_2$
from (\ref{betatwo}) and then $\zeta_1$ from (\ref{secondwave}).

Thereafter we can proceed with the computation of the full
second order. This scheme generalizes to arbitrary order of
perturbation theory. Assume that we have computed $V^{(m)}(\phi)$
for all orders $1\leq m\leq n-1$ except for the wave function
renormalization $\zeta_{n-1}$ at order $n-1$. Then the scaling
equation
\bea
&& b_n \PF V^{(1)}(\psi)=-n\PF\left\langle V^{(1)};V^{(n-1)}
\right\rangle^T_{\Gamma,\SG\psi}+
\label{betanew}\\
&&\sum_{m=2}^{n-1}
\frac{(-1)^{m+1}}{m!}
\sum_{\begin{array}{c}l_1,\dots,l_m\in\N\\
l_1+\cdots+l_m=n\end{array}}\left(\begin{array}{c}
n\\l_1\cdots l_{m-1}\end{array}\right)
\PF\left\langle V^{(l_1)};\cdots;V^{(l_m)}
\right\rangle^T_{\Gamma,\SG\psi}
\nonumber
\eea
allows us to determine the $n$'th order coefficient $b_n$
of the $\beta$-function. Analogous to the second order case
one does not find $\zeta_{n-1}$ on the right side of
(\ref{betanew}). To make the mechanism explicit let us split
off the wave function term from the effective interaction and
write
\beq
V^{(m)}(\psi)=\zeta_m \OBS_{2,2}(\psi)+
W^{(m)}(\psi).
\eeq
Then by the same reasons as above it follows that
\beq
\PF\left\langle V^{(1)};V^{(n-1)}\right\rangle_{\Gamma,
\SG\psi}=
\PF\left\langle W^{(1)};W^{(n-1)}\right\rangle_{\Gamma,
\SG\psi}
\eeq
is independent of $\zeta_{n-1}$. But all other contributions
in (\ref{betanew}) stem from lower orders only. The final
piece of gymnastics is to find $\zeta_{n-1}$ from
\bea
&&
\frac{n(n-1)}2(b_1)^{n-2} b_2\PT V^{(n-1)}(\psi)=
-b_n\PT V^{(1)}(\psi)-nb_1b_{n-1}\PT V^{(2)}(\psi)-
\nonumber\\&&\quad
\sum_{m=2}^{n-1}\frac1{m!}
\sum_{\begin{array}{c}l_1,\dots,l_m\in\N\\
l_1+\cdots+l_m=n\end{array}}\left(\begin{array}{c}
n\\l_1\cdots l_{m-1}\end{array}\right)
\biggl(b_{l_1}\cdots b_{l_m} V^{(m)}(\psi)+
\nonumber\\&&\quad
(-1)^m\left\langle V^{(l_1)};\cdots;V^{(l_m)}
\right\rangle^T_{\Gamma,\SG\psi}\biggr)-
n\PT\left\langle V^{(1)};V^{(n-1)}
\right\rangle^T_{\Gamma,\SG\psi}+
\label{finalwork}\\&&\quad
(-1)^{(n+1)}\frac{n(n-1)}2\left\langle V^{(1)};\cdots;
V^{(1)};V^{(2)}\right\rangle^T_{\Gamma,\SG\psi}+
\quad(-1)^n\left\langle V^{(1)};\cdots;V^{(1)}
\right\rangle^T_{\Gamma,\SG\psi}.
\nonumber
\eea
Again the right side does not depend on the order $n-1$
wave function renormalization constant $\zeta_{n-1}$ because
\beq
\PT\left\langle V^{(1)};V^{(n-1)}\right\rangle_{\Gamma,
\SG\psi}=
\PT\left\langle W^{(1)};W^{(n-1)}\right\rangle_{\Gamma,
\SG\psi}.
\label{vanish3}
\eeq
Notice that the wave function renormalization constant is
hidden notationally on the left side of (\ref{finalwork})
in
\beq
\PT V^{(n-1)}(\psi)=\zeta_{n-1}\OBS_{2,2}(\psi).
\eeq
Notice also that it comes together with pre-factors which do
not vanish. The iterative scheme is now complete. In summary
it goes as follows. To order $n$ we first compute $b_n$ from
(\ref{betanew}), then $\zeta_{n-1}$ from (\ref{finalwork}),
and thereafter the effective interaction $V^{(n)}(\psi)$ except
for $\zeta_n$. Then we iterate the computation in the next order.
%
%====================================================================
\section{Estimates}

In the iteration the lower order vertices are convoluted
with hard and soft propagators to form the next order
inhomogeneous scaling terms. In this section we present an
estimate which shows that the integrals in this process
converge. Consequently the expansion is finite
to every order of perturbation theory. In a nutshell the
expansion is finite, because the induction step involves
perturbation theory with cutoff propagators and smooth vertices
which do not grow too fast at large momenta. The fluctuation
propagator has two sided cutoffs. The estimates
\bea
\|\widehat{\Gamma}\|_{\infty,-2\epsilon}&=&
\sup_{p\in\R^D}\left\{\left\vert\widehat{\Gamma}(p)
\right\vert e^{2\epsilon |p|}\right\} <\infty, \\
\|\widehat{\Gamma}\|_{1,-2\epsilon}&=&
\int \frac{{\rm d}^Dp}{(2\pi)^D}
\left\vert\widehat{v}(p)\right\vert
\;e^{2\epsilon |p|}<\infty,
\eea
for $\epsilon$ sufficiently small, should therefore cause
no surprise. We save a small amount of exponential decrease
to bound the large momentum growth of the vertices. It is
then fairly obvious that a perturbation theory with this
fluctuation propagator and $L_{\infty,\epsilon}$-vertices
is finite. The normal ordering propagator has an ultraviolet
unit cutoff but no infrared cutoff. It satisfies
\beq
\|\widehat{v}\|_{1,-2\epsilon}<\infty.
\eeq
But this integrability in dimensions larger than two is
sufficient to bound loop integrals with normal ordering
propagator. Fortunately they never occur at external legs,
where they would spoil the Taylor coefficients.

Recall that the induction step to order $n$ involves
the computation of $b_n$, $\zeta_{n-1}$, and thereafter
$V^{(n)}(\psi)$ (except for $\zeta_n$). We present
the argument for finiteness of $V^{(n)}(\psi)$ to
some detail. Finiteness of the coefficients then
follows by the same reasoning. Let us assume bounds on
$b_n$, $\zeta_{n-1}$, $V^{(n-1)}(\psi)$, and respective
lower orders. The right side of the recursion equations
(\ref{recursion}) consists of two contributions,
the {\it dynamical} contribution
\beq
\sum_{m=2}^{n}\frac{(-1)^m}{m!}
\sum_{\begin{array}{c}l_1,\dots,l_m\in\N\\
l_1+\cdots+l_m=n\end{array}}\left(\begin{array}{c}
n\\l_1\cdots l_{m-1}\end{array}\right)
\left\langle V^{(l_1)};\cdots;V^{(l_m)}
\right\rangle^T_{\Gamma,\SG\psi},
\label{firstcon}
\eeq
from the fluctuation integral, and the {\it kinematical}
contribution from the flow of the coupling,
\beq
\sum_{m=1}^{n-1}\frac1{m!}
\sum_{\begin{array}{c}l_1,\dots,l_m\in\N\\
l_1+\cdots+l_m=n\end{array}}\left(\begin{array}{c}
n\\l_1\cdots l_{m-1}\end{array}\right)
b_{l_1}\cdots b_{l_m} V^{(m)}(\psi).
\eeq
The second one immediately inherits a bound from the
induction hypothesis on the lower orders. Recall that
it arises from a re-organisation of the interaction as a
polynomial of the bare coupling into a polynomial of the
effective coupling. This re-organisation is finite because
the lower order coefficients of the $\beta$-function are
finite. To estimate the (\ref{firstcon}), we break it down
to a sum of Feynman amplitudes. Fortunately we do not
need more information besides that it can be written as
a sum of finitely many amplitudes with certain properties.
Each of the lower order interactions is a sum of vertices
$V^{(l)}(\psi)=\sum_{k=1}^{l+1}V_{2k}^{(l)}(\psi)$ of
the form
\beq
V^{(l)}_{2k}(\psi)=
\frac1{(2k)!}\int
{\rm d}^Dx_1\cdots{\rm d}^Dx_{2k}
:\psi(x_1)\cdots\psi(x_{2k}):
V^{(l)}_{2k}(x_1,\dots,x_{2k}).
\label{kernels}
\eeq
We estimate individually each contribution in the truncated
expectation value
\beq
\left\langle V^{(l_1)};\dots;V^{(l_m)}
\right\rangle^T_{\Gamma,\SG\psi}=
\sum_{k_1=1}^{l_1+1}\cdots\sum_{k_m=1}^{l_m+1}
\left\langle V^{(l_1)}_{2k_1};\dots;V^{(l_m)}_{2k_m}
\right\rangle^T_{\Gamma,\SG\psi}
\eeq
for the sake of notational economy. Insert the expression
(\ref{kernels}). Then the task becomes an estimate for
\bea
\left\langle V^{(l_1)}_{2k_1};\cdots;V^{(l_m)}_{2k_m}
\right\rangle^T_{\Gamma,\SG\psi}&=&
\int\left(\prod_{j=1}^m\prod_{i=1}^{2k_j}
{\rm d}^Dx_{j,i}\right)\prod_{j=1}^m
V^{(l_j)}_{2k_j}(x_{j,1},\dots,x_{j,2k_j})
\nonumber \\ & & \quad
\left\langle\prod_{j=1}^m\left[:\prod_{i=1}^{2k_j}
\phi(x_{j,i}):;\right]\right\rangle^T_{\Gamma,\SG\psi}.
\label{task}
\eea
The truncated expectation value in (\ref{task}) contains
a product of clusters, each cluster being a normal ordered
products of fields. The index $j$ defines a colouring of the
clusters. Eq. (\ref{task}) is computed in two steps. Step one
is
\bea
\left\langle\prod_{j=1}^m\left[:\prod_{i=1}^{2k_j}
\phi(x_{j,i}):_v;\right]\right\rangle^T_{\Gamma,\SG\psi}
&=&\sum_{I_1\subset\{1,\dots,2k_1\}}\cdots
\sum_{I_m\subset\{1,\dots,2k_m\}}
\prod_{j=1}^m :\prod_{i\in I_j}\SG\psi(x_{j,i}):_{v-\Gamma}
\nonumber \\ & &\quad
\left\langle\prod_{j=1}^m\left[:
\prod_{i\in\{1,\dots,2k_j\}\setminus I_j}
\zeta(x_{j,i}):_\Gamma;\right]\right\rangle^T_{\Gamma}.
\eea
The truncated expectation value is here a standard one
with mean zero. We sum over all subsets $I_j$ of
$\{1,\dots,2k_j\}$. The interpretation is that for each
field $\phi(x_{j,i})$ belonging to cluster $j$ we decide
whether it be a rescaled background field $\SG\psi(x_{j,i})$
or a fluctuation field $\zeta(x_{j,i})$. The truncated
expectation value is zero unless both the total number
of fluctuation fields is even, and each cluster contains
at least one fluctuation field. It is evaluated with Wick's
theorem as a sum over pairings. The normal ordering
forbids pairings within clusters. Step two is a re-normal
ordering of the background fields. This is done with the
help of
\bea
\prod_{j=1}^m:\prod_{i\in I_j}\SG\psi(x_{j,i}):_{v-\Gamma}
&=&\sum_{J_1\subseteq I_1}\cdots\sum_{J_m\subseteq I_m}
:\prod_{j=1}^m\prod_{i\in J_j} \SG\psi(x_{j,i}):_{v-\Gamma}
\nonumber \\ & & \quad
\left\langle \prod_{j=1}^m :\prod_{i\in J_j^c}
\phi (x_{j,i}):_{-v} \right\rangle_{-v}.
\label{renormal}
\eea
The Gaussian expectation value is here un-truncated and
with negative covariance. It is defined by Wick's rule
as a sum of all pairings. The number of fields being
contracted is again required to be even. The expectation
value is zero for an odd number of fields. The sums in
(\ref{renormal}) go over all subsets $J_j$ of $I_j$
including the empty set and $I_j$ itself. Their meaning
is that for each field in cluster $j$ we decide whether it
be a truly external field or contracted with a field
in another cluster $j^\prime\neq j$. We can think
of this process as a generation of additional loops with
normal ordering covariance in every term generated by
the fluctuation integral. Having done both the truncated
fluctuation integral and the un-truncated normal ordering
"integral" we obtain
\bea
\left\langle V^{(l_1)};\cdots;V^{(l_m)}
\right\rangle^T_{\Gamma,\SG\psi}&=&
\sum_{p=1}^{\sum_{j=1}^m(k_j-1)+1}
\frac1{(2p)!}\int{\rm d}^Dy_1\cdots{\rm d}^Dy_{2p}
\nonumber \\ & & \quad
L^{p(D+2)} K^{(l_1,\cdots,l_m)}_{2p}(Ly_1,\dots,Ly_{2p})
\eea
with kernels given by
\bea
& &K^{(l_1,\cdots,l_m)}_{2p}(y_1,\dots,y_{2p})=
(2p)!\sum_{I_1\subset\{1,\dots,2k_1\}}\cdots
\sum_{I_m\subset\{1,\dots,2k_m\}}
\sum_{J_1\subseteq I_1}\cdots\sum_{J_m\subseteq I_m}
\nonumber \\ & & \quad
\delta_{|J_1|+\cdots+|J_m|,2p}
\int\left(\prod_{j=1}^m
\prod_{i\in\{1,\dots,2k_j\}\setminus J_j}
{\rm d}^Dx_{j,i}\right)
\prod_{j=1}^m\left(\frac1{(2k_j)!}
V^{(l_j)}_{2k_j}(x_{j,1}),\dots,x_{j,2k_j})\right)
\nonumber\\ & &\quad
\left\langle\prod_{j=1}^m
\left[:\prod_{i\in\{1,\dots,2k_j\}\setminus I_j}
\zeta(x_{j,i}):_\Gamma;\right]
\right\rangle^T_{\Gamma}
\left\langle\prod_{j=1}^m:\prod_{i\in I_j\setminus J_j}
\phi(x_{j,i}):_{-v}\right\rangle_{-v}.
\label{diag}
\eea
The external points on the right side are here understood
to be renamed. There order is of no importance since all
kernels are symmetric. The terms on the right hand side
depend on the cardinalities $|I_j|$ and $|J_j|$ only.
We are now ready to do the estimates. The right hand side
of (\ref{diag}) is a sum of terms coming from writing out
explicitely the expectation values in terms of contractions,
the Feynman amplitudes. Their precise form is of no importance
here. We restrict our attention to the number of lines and
loops and combinatoric factors and find
\bea
L&=&\frac12\sum_{j=1}^m(2k_j-|J_j|), \\
L_H&=&\frac12\sum_{j=1}^m(2k_j-|I_j|), \\
L_H^{tree}&=&m-1, \\
L_S&=&\frac12\sum_{j=1}^m(|I_j|-|J_j|).
\eea
Here $L$ is the total number of lines, that is, factors
of propagators. $L_H$ is the number of hard lines of
fluctuation propagators coming from the truncated
expectation value. $L_S$ is the number of soft lines
from re-normal ordering. Each contraction is cluster
connected in terms of hard lines. Therefore each
contraction contains a tree of hard lines. This number
of tree hard lines is $L_H^{tree}$. The difference
of $L$ and $L_H^{tree}$ is the number of loop integrals.
We are ready to do the estimate. Put
$L_{1,-2\epsilon}$-norms on all loop lines of hard and soft
propagators. Put $L_{\infty,-2\epsilon}$-norms on the
hard tree lines. Put $L_{\infty,\epsilon}$-norms on all
vertices. Each of the terms on the right side of
(\ref{diag}) being bounded we find
\bea
& &\| \widehat{K}^{(l_1,\dots,l_m)}_{2p}
\|_{\infty,\epsilon} \leq
(2p)!\sum_{|I_1|=0}^{2k_j-1}\cdots\sum_{|I_m|=1}^{2k_m-1}
\sum_{|J_1|=0}^{|I_1|}\cdots\sum_{|J_m|=0}^{|I_m|}
\delta_{|J_1|+\cdots+|J_m|,2p}
\nonumber \\ & & \quad
\prod_{j=1}^m\left(
\left(\begin{array}{c} 2k_j\\ |I_j|\end{array}\right)
\left(\begin{array}{c} |I_j|\\ |J_j|\end{array}\right)
\frac1{2k_j)!}
\| \widehat{V}^{(l_j)}_{2k_j}\|_{\infty,\epsilon}\right)
\| \widehat{\Gamma}\|_{\infty,-2\epsilon}^{L_H^{tree}}
\|\widehat{\Gamma}\|_{1,-2\epsilon}^{L_H-L_H^{tree}}
\|\widehat{v}\|_{1,-2\epsilon}^{L_S}
\nonumber \\ & & \quad
(2L_H-1)!! (2L_S-1)!!.
\eea
More sophisticated estimates will be presented elsewhere. Hence
$\| \widehat{K}^{(l_1,\dots,l_m)}_{2p}\|_{\infty,\epsilon}$
is finite in momentum space. Furthermore all loop integrals
converge, and the result is smooth in the external momenta.
It can in particular be Taylor expanded to any desired order.
But then we are done: the right hand side of the recursion
relation is smooth and can therefore be solved as above.
Thus the expansion is indeed finite to every order of
perturbation theory.
%
%====================================================================
\section{Second order}

The second order scaling equation (\ref{secord}) involves
a truncated expectation value of two first order interactions.
The first order interaction is given by eq. (\ref{newone}),
including a $\phi^4$-vertex and a wave function term.
The first order wave function constant will come out of
this second order computation. The expectation value is
\beq
\left\langle V^{(1)};V^{(1)}\right\rangle^T_{\Gamma,\SG\psi}=
\left\langle\OBS_{4,0};\OBS_{4,0}\right\rangle^T_{\Gamma,\SG\psi}+
2\zeta_1
\left\langle\OBS_{4,0};\OBS_{2,2}\right\rangle^T_{\Gamma,\SG\psi}+
\zeta_1^2
\left\langle\OBS_{2,2};\OBS_{2,2}\right\rangle^T_{\Gamma,\SG\psi}.
\eeq
Each contribution is best taken care of separately. The computation
consists of three steps. The first step is the fluctuation integral.
It generates contractions between the vertices and shifts the
normal ordering covariance. The second step is the rescaling of
the external field. It restores the invariant normal ordering
covariance and rescales the kernels of the effective interactions.
The third step is a re-normal ordering of the result. Re-normal
ordering creates additional loops with normal ordering propagators.
The final result is
\bea
\left\langle\OBS_{4,0};\OBS_{4,0}\right\rangle^T_{\Gamma,\SG\psi}
&=&
\frac 12 \int{\rm d}^Dx\int{\rm d}^Dy :\psi(x)\psi(y):
L^{2+D}\biggl(\frac 13\Gamma(Lx-Ly)^3+
\nonumber\\ & & \quad
\Gamma(Lx-Ly)^2 u(Lx-Ly)+\Gamma(Lx-Ly) u(Lx-Ly)^2\biggr)+
\nonumber\\ & &
\frac 1{4!} \int{\rm d}^Dx\int{\rm d}^Dy :\psi(x)^2\psi(y)^2:
L^4 \biggl(3\Gamma(Lx-Ly)^2+
\nonumber\\ & & \quad
6\Gamma(Lx-Ly) u(Lx-Ly)\biggr)+
\nonumber\\ & &
\frac 1{6!} \int{\rm d}^Dx\int{\rm d}^Dy :\psi(x)^3\psi(y)^3:
20L^{6-D}\Gamma(Lx-Ly),
\\
2\left\langle\OBS_{4,0};\OBS_{2,2}\right\rangle^T_{\Gamma,\SG\psi}
&=&
\frac 12 \int{\rm d}^Dx :\psi(x)^2:
L^2 \biggl(
\int{\rm d}^Dy \Gamma(y)
\frac{-\partial^2}{\partial y^2} \Gamma(y)+
\nonumber\\ & & \quad
2\int{\rm d}^Dy
v(y)\frac{-\partial^2}{\partial y^2} \Gamma(y) \biggr)+
\nonumber\\ & &
\frac 1{4!} \int{\rm d}^Dx\int{\rm d}^Dy :\psi(x)^3\psi(y):
8L^2\frac{-\partial^2}{\partial y^2} \Gamma(Lx-Ly),
\\
\left\langle\OBS_{2,2};\OBS_{2,2}\right\rangle^T_{\Gamma,\SG\psi}
&=&
\frac 12 \int{\rm d}^Dx\int{\rm d}^Dy :\psi(x)\psi(y):
2L^{-2+D}
\frac{-\partial^2}{\partial x^2}
\frac{-\partial^2}{\partial y^2}
\Gamma(Lx-Ly).
\eea
Here $u(x)=L^{-2+D}v(L^{-1}x)$ denotes a rescaled normal
ordering covariance. The partial derivatives are meant to
act on rescaled fluctuation propagators.
The relevant part can be extracted by Taylor expansion of the
Fourier transformed kernels. The inhomogeneous terms of the
second order scaling equation are given by
\bea
&&\widehat{K}^{(2)}_2(p_1,p_2)=
L^{-2(4-D)}\biggl(b_2\zeta_1 p_1^2+
L^2\zeta_1(c_1+2c_2)+L^{2-D}\zeta_1^2\widehat{\Gamma}(L^{-1}p_1)+
\nonumber\\ &&\qquad
\frac{L^2}3\widehat{\Gamma}\star\widehat{\Gamma}\star\widehat{\Gamma}
(L^{-1}p_1)+
L^2\widehat{\Gamma}\star\widehat{\Gamma}\star\widehat{u}
(L^{-1}p_1)+
L^2\widehat{\Gamma}\star\widehat{u}\star\widehat{u}
(L^{-1}p_1)\biggr),
\\
&&\widehat{K}^{(2)}_4(p_1,\dots,p_4)=
L^{-2(4-D)}\biggl(b_2+8L^{2-D}p_1^2\widehat{\Gamma}(L^{-1}p_1)+
\nonumber\\ &&\qquad
3L^{4-D}\widehat{\Gamma}\star\widehat{\Gamma}
(L^{-1}p_1+L^{-1}p_2)+
6L^{4-D}\widehat{\Gamma}\star\widehat{u}
(L^{-1}p_1+L^{-1}p_2)\biggr)
\\
&&\widehat{K}^{(2)}_6(p_1,\dots,p_6)=
L^{-2(4-D)}\biggl(20 L^{6-2D}
\widehat{\Gamma}(L^{-1}p_1+L^{-1}p_2+L^{-1}p_3)\biggr),
\eea
symmetrized in their momenta. The sum of momenta is constrained
to zero. Here $\star$ denotes convolution in momentum space
(times $(2\pi)^{-D}$). The constants stand for the convergent
loop integrals
\bea
c_1&=&\int\frac{{\rm d}^Dp}{(2\pi)^D}p^2\widehat{\Gamma}(p)^2,
\\
c_2&=&\int\frac{{\rm d}^Dp}{(2\pi)^D}p^2\widehat{\Gamma}(p)
\widehat{u}(p).
\eea
The momentum space convolutions are conveniently computed
with the help of the parameter representations
\bea
\widehat{\Gamma}(p)&=&
\int_1^{L^2}{\rm d}\alpha e^{-\alpha p^2}\\
\widehat{u}(p)&=&
\int_{L^2}^{\infty}{\rm d}\alpha e^{-\alpha p^2}.
\eea
The momentum space integrals become Gaussian and can be
reduced to the two formulas
\bea
\int\frac{{\rm d}^Dp}{(2\pi)^D}
e^{-\alpha_1 p^2-\alpha_2 (p-q)^2}&=&
(4\pi)^{\frac{-D}2}(\alpha_1+\alpha_2)^{\frac{-D}{2}}
e^{\frac{-\alpha_1\alpha_2}{\alpha_1+\alpha_2} q^2},
\\
\int\frac{{\rm d}^Dp_1}{(2\pi)^D}
\int\frac{{\rm d}^Dp_2}{(2\pi)^D}
e^{-\alpha_1 p_1^2-\alpha_2 p_2^2-
\alpha_3 (p_1+p_2-q)^2}&=&
(4\pi)^{-D}(\alpha_1\alpha_2+\alpha_1\alpha_3+
\alpha_2\alpha_3)^{\frac{-D}{2}}
\nonumber \\ & &
e^{\frac{-\alpha_1\alpha_2\alpha_3}
{\alpha_1\alpha_2+\alpha_1\alpha_3+\alpha_2\alpha_3}
q^2}.
\eea
A welcome feature of this parameter representation is that the
result depends only on the external momentum squared. The
scaling equation for the coefficient $b_2$ in the $\beta$-function
follows from evaluation at zero momentum. It is given by
\beq
b_2=-L^{4-D}\left(3\widehat{\Gamma}\star\widehat{\Gamma}(0)+
6 \widehat{\Gamma}\star\widehat{u}(0)\right).
\eeq
Its four dimensional value with exponential cutoff is computed
to
\beq
b_2=\frac{-6\log (L)}{(4\pi)^2}.
\eeq
It follows that the second order flow on the renormalized
trajectory in four dimensions is given by
\beq
\beta(g)=g-\frac{3\log (L)}{(4\pi)^2}g^2+
O\left(g^3\right).
\eeq
It follows that the model is asymptotically free in the
infrared limit at weak coupling. The scaling equation for
the first order wave function constant is found to be
\beq
\zeta_1=\frac{-1}{b_2}
\frac{\partial}{\partial (p^2)}
\left(\frac 13
\widehat{\Gamma}\star\widehat{\Gamma}\star\widehat{\Gamma}(p)+
\widehat{\Gamma}\star\widehat{\Gamma}\star\widehat{u}(p)+
\widehat{\Gamma}\star\widehat{u}\star\widehat{u}(p)
\right)\Biggr\vert_{p^2=0}.
\eeq
The result in four dimensions is
\beq
\zeta_1=\frac {-1}{18(4\pi)^2}.
\eeq
There is one relevant coordinate left at second order, the
quadratic kernel at zero momentum. It is directly determined
by the scaling equation
\bea
\left(L^{2-2(4-D)}-1\right)\widehat{V}^{(2)}_2(0,0)&=&
L^{-2(4-D)}\biggl( L^2\zeta_1(c_1+2c_2)+
\nonumber\\ & &
\frac{L^2}3
\widehat{\Gamma}\star\widehat{\Gamma}\star\widehat{\Gamma}(0)+
L^2 \widehat{\Gamma}\star\widehat{\Gamma}\star\widehat{u}(0)+
L^2 \widehat{\Gamma}\star\widehat{u}\star\widehat{u}(0)
\biggr).
\eea
Doing these integrals as a last exercise the effective mass
constant at second order in four dimensions turns out as
\beq
\widehat{V}^{(2)}_2(0,0)=
\frac {1}{(4\pi)^4}
\left(2\log (2)-\log (3)-\frac 1{36}\right).
\eeq
This completes the computation of the second order renormalization
constants. The full second order kernels are then given by
summed scaled parameter integrals. The explicit computation of
this irrelevant part will not be pursued further here.
If this second order approximation is for instance used in
a numerical simulation as renormalization group improved action,
it is reasonable to approximate it by a momentum space Taylor
expansion to some chosen degree of irrelevancy. Then this
computation supplies the non-irrelevant part.
%
%====================================================================
\section{Conclusions}

The standard renormalization scheme departs from a bare
action. The renormalized trajectory is reached upon infinite
iteration of block spin transformations. A further block
spin transformation does no harm to a renormalized action.
It merely generates a renormalized renormalization
group flow on the renormalized trajectory. In view of this
renormalization group flow on the renormalized trajectory we
speak of running couplings. We should mention that the behavior
of a renormalized action under the field theoretic
renormalization group was used by Callan \cite{C76} in his
proof of the BPHZ theorem, a polished version of which was
given by Lesniewski \cite{L83}. By universality the
trajectory can be approached from a variety of bare actions.
The notion of bare action is not unique. Consider for instance
again the ultraviolet limit of $\phi^4$-theory in four
dimensions (say at negative coupling). The bare action is required
to converge to the stable manifold of the trivial fixed point.
But this stable manifold of irrelevant couplings is infinite
dimensional. The approach presented here is free of bare
ambiguities. It is designed upon equations for the renormalized
action.

The renormalized trajectory can be viewed itself as
a means to investigate ultraviolet and infrared limit of
a Euclidean field theory. It is ideally suited to perform
an infinite number of block spin transformations. A single
block spin step on the trajectory translates to
a (generally non-linear) transformation on the low
dimensional space of running couplings. As a finite
dimensional dynamical system it is comparatively easy to
analyze. Consider for instance the $\phi^4$-trajectory
in the formalism above. There a block spin transformation
becomes a transformation of a running coupling $g$ in terms
of a step $\beta$-function $\beta (g)=b_1 g+b_2 g^2+
O(g^3)$ with coefficients $b_1=L^{4-D}$ and $b_2 <0$.
Here $L>1$ is the block scale and $D$ is the Euclidean dimension.
For $D=4$ it follows that the flow on the $\phi^4$-trajectory
is not asymptotically free at weak positive coupling in the
ultraviolet limit. The running coupling shrinks under a block
spin transformation. An infinite number of block spin
transformations require either the renormalized coupling to
approach zero or the bare coupling (as is believed) to tend
to infinity. Perturbation theory is in this sense renormalization
group inconsistent for the ultraviolet limit in four dimensions.
We remark that another non-trivial renormalization group
fixed point on the renormalized trajectory would allow to
perform the ultraviolet limit. It would albeit suffice to
require an infinite number of inverse block spin transformations
for the running coupling to diverge. Either way requires control
of the renormalized trajectory outside a vicinity of the
trivial fixed point. This question is believed to be
settled in favour of triviality by Aizenmann and Fr\"ohlich.
See \cite{FFS92} and references therein. However even in
this situation of an inconsistency of perturbation theory
in the ultraviolet limit perturbation theory for the
$\phi^4$-trajectory at weak couplings is a well posed problem.

The computational scheme presented here is very flexible.
It applies to any form of block spin transformations
in the vicinity of a trivial fixed point.
The expansion is iterative and built upon a fairly
simple recursion relation. One can therefore expect it
to be useful for accurate bounds on renormalized series.
For instance tree decay of interactions is an immediate
consequence. However the $L_\infty$-bound is already sharper
than the estimates of Polchinski \cite{P84}, and the
improvements by Keller, Kopper, and Salmhofer \cite{KKS90},
and by Hurd \cite{H89}. It does not lead to logarithms in a
renormalization scale. Other sophisticated expansion
technology can be found in Rivasseau's textbook
\cite{R91} together with a set of references to
the Paris school of renormalization. In particular
Rivasseau's effective expansion seems to be related
to the one considered here.

The question whether our approach makes sense
beyond perturbation theory is a very interesting one.
Whether renormalization group invariance plus certain
intial conditions determine a renormalized trajectory
beyond perturbation theory we do not know.
It is certainly the case where the perturbation
expansion converges. Last not least it would be
highly desirable to develop iterative techniques for
unstable manifolds of nontrivial fixed points.
A clear presentation of the weak coupling case around
the trivial fixed point could be part of this way.
%
%====================================================================

%
\end{document}